\documentclass[aps,twocolumn,showpacs,showkeys,nofootinbib]{revtex4}
\usepackage{epsfig}
\usepackage{amsmath}
\usepackage{amsfonts}
\usepackage{amssymb}
\usepackage{graphicx}
\usepackage{colordvi}
\usepackage[T1]{fontenc}

\begin{document}

\title{Numerical Analysis of Galactic Rotation Curves}

\author{G. Scelza\footnote{e - mail address: lucasce73@gmail.com}, A. Stabile$^1$\footnote{e -
mail address: arturo.stabile@gmail.com}}

\affiliation{$^1$Dipartimento di Ingegneria, Universit\`{a} del Sannio,
Palazzo Dell'Aquila Bosco Lucarelli, Corso Garibaldi, 107 - 82100,
Benevento, Italy}

\begin{abstract}
In this paper we present the discussion on the salient points of the computational analysis that are at the basis of the paper \emph{Rotation Curves of Galaxies by Fourth Order Gravity} \citep{StSc}. In fact in this paper any galactic component (bulge, disk and Dark matter component) required an onerous numerical computation since the Gauss theorem is not applicable in the Fourth Order Gravity. The computational and data analysis have been made with the software Mathematica$^\circledR$.
\end{abstract}

\keywords{Galactic rotation curves, numerical analysis.}
\maketitle

\section{Introduction}

Today the Universe appears spatially flat undergoing an accelerated expansion. There are many measurements proving this pictures \cite{riess, ast, clo, ber, spe, carrol, staro, sahini}. According to the successful cosmological model \cite{carrol, staro, sahini}, there are two main ingredients in this scenario, namely Dark Matter (DM) and the cosmological constant $\Lambda$ (Dark Energy). On the galactic scales, the evolution is driven by the usual Newtonian gravitational potential, but it needs hypothesizing the existence of DM to obtain a good experimental agreement. A good model for the galactic distribution of DM, in the framework of General Relativity (GR), is the Navarro-Frenk-White model (NFW model) \cite{navarro}.

However in recent years, the effort to give a physical explanation to the cosmic acceleration has attracted an amount of interest in so called Fourth Order Gravity (FOG), and particularly the $f(R)$-Gravity, where $f$ is a generic function of Ricci scalar $R$. These alternative models have been considered as a viable mechanism to explain the cosmic acceleration. Apart the cosmological dynamics, a systematic analysis of such theories were performed at short scale and in the low energy limit \cite{olmo1, olmo2, olmo3, Damour:Esposito-Farese:1992, clifton, odintsov, newtonian_limit_fR, PRD, stelle, schm, FOG, Stabile_Capozziello}.

In particular the paper \emph{Most General Fourth Order Theory of Gravity at Low Energy} \cite{FOG} analyzed the gravitational potential, induced by a $f(X,Y,Z)$-Gravity, where for sake of simplicity we set $X\,=\,R$, $Y\,=\,R^{\alpha\beta}R_{\alpha\beta}$ and $Z\,=\,R^{\alpha\beta\gamma\delta}R_{\alpha\beta\gamma\delta}$, generalizing the Hilbert Einstein lagrangian. The added quantities are the Ricci tensor $R_{\mu\nu}$ and the Riemann tensor $R_{\mu\nu\alpha\beta}$. As astrophysical application the modified potential has been used to build the rotation curves for the Milky Way and NGC 3198 \cite{StSc}. In this paper any galactic component (bulge, disk and DM component) required an onerous numerical computation since the Gauss theorem is not applicable in the FOG. The aim of the present paper is to point out the fundamental topics of the adopted strategy using the software \emph{Mathematica}$^\circledR$.

Our analysis is then organized as follows. In section \ref{theory} we report the fundamental topics of the fourth order gravity: the field equations and their newtonian approximation, the solution for the gravitational potential and the mathematical models for the galactic componets. In section \ref{computation} we build we build the code for the numerical simulation and in section \ref{data} there is the data fit between our theoretical curves and the data of the rotation curve of the Milky Way and the galaxy NGC 3190. Finally in section \ref{conclusions} we report the conclusions.

\section{The galactic rotation curves in the framework of $f(X,Y,Z)$-Gravity}\label{theory}

Let us start with a general class of FOG given by the action

\begin{eqnarray}\label{FOGaction}
\mathcal{A}\,=\,\int d^{4}x\sqrt{-g}\biggl[f(X,Y,Z)+\mathcal{X}\mathcal{L}_m\biggr]
\end{eqnarray}
where $f$ is an unspecified function of curvature invariants.
The term $\mathcal{L}_m$ is the minimally coupled ordinary matter
contribution. In the metric approach, the field equations are
obtained by varying (\ref{FOGaction}) with respect to
$g_{\mu\nu}$. We get

\begin{widetext}
\begin{eqnarray}\label{fieldequationFOG}
\begin{array}{ll}
f_XR_{\mu\nu}-\frac{f}{2}g_{\mu\nu}-f_{X;\mu\nu}+g_{\mu\nu}\Box
f_X+2f_Y{R_\mu}^\alpha
R_{\alpha\nu}-2[f_Y{R^\alpha}_{(\mu}]_{;\nu)\alpha}\\\\\,\,\,\,\,\,\,\,\,\,\,\,\,\,\,\,\,\,\,\,\,\,\,\,\,\,\,\,\,\,\,\,\,\,
\,\,\,\,\,\,\,\,\,\,\,\,\,\,\,+\Box[f_YR_{\mu\nu}]+[f_YR_{\alpha\beta}]^{;\alpha\beta}g_{\mu\nu}
+2f_ZR_{\mu\alpha
\beta\gamma}{R_{\nu}}^{\alpha\beta\gamma}-4[f_Z{{R_\mu}^{\alpha\beta}}_\nu]_{;\alpha\beta}\,=\,
\mathcal{X}\,T_{\mu\nu}
\\\\
f_XR+2f_YR_{\alpha\beta}R^{\alpha\beta}+2f_ZR_{\alpha\beta\gamma\delta}
R^{\alpha\beta\gamma\delta}-2f+\Box[3
f_X+f_YR]+2[(f_Y+2f_Z)R^{\alpha\beta}]_{;\alpha\beta}\,=\,\mathcal{X}\,T
\end{array}
\end{eqnarray}
\end{widetext}
where $f_X\,=\,\frac{\partial f}{\partial X}$, $f_Y\,=\,\frac{\partial f}{\partial Y}$,
$f_Z\,=\,\frac{\partial f}{\partial Z}$, $\Box={{}_{;\sigma}}^{;\sigma}$ and
$\mathcal{X}\,=\,8\pi G$. $T_{\mu\nu}\,=\,-\frac{1}{\sqrt{-g}}\frac{\delta(\sqrt{-g}\mathcal{L}_m)}{\delta g^{\mu\nu}}$ is the energy-momentum tensor of matter and $T$ is
its trace. The second line of (\ref{fieldequationFOG}) is the
trace of the first one.

In the case of weak field and slow motion we consider the field
equation in the so called Newtonian limit of theory. For our aim
we can consider the metric tensor approximated as follows (for
details, see \cite{newtonian_limit_fR, rew, landau, PRD})

\begin{eqnarray}\label{metric_tensor_PPN}
  g_{\mu\nu}\,=\,\begin{pmatrix}
  1+2\,\Phi(t,\mathbf{x})& 0 \\
  \\
  0 & -\delta_{ij}\end{pmatrix}
\end{eqnarray}
where $\Phi$ is the gravitational potentials and $\delta_{ij}$ is
the Kronecker delta. The set of coordinates adopted is $x^\mu\,=\,(t,x^1,x^2,x^3)\,=\,(t,\mathbf{x})$. By
introducing the quantities

\begin{eqnarray}\label{mass_definition}
\begin{array}{ll}
{m_1}^2\,\doteq\,-\frac{f_X(0)}{3f_{XX}(0)+2f_Y(0)+2f_Z(0)}\\\\
{m_2}^2\,\doteq\,\frac{f_X(0)}{f_Y(0)+4f_Z(0)}
\end{array}
\end{eqnarray}
we get three differential equations for the curvature invariant $X$ and the gravitational potentials $\Phi$, $\Psi$

\begin{eqnarray}\label{NL-field-equation_2}
\begin{array}{ll}
(\triangle-{m_2}^2)\triangle\Phi+\biggl[\frac{{m_2}^2}{2}-\frac{{m_1}^2+2{m_2}^2}{6{m_1}^2}\triangle\biggr]
X\,=\,-{m_2}^2\mathcal{X}\,\rho\\\\
(\triangle-{m_1}^2)X\,=\,{m_1}^2\mathcal{X}\,\rho
\end{array}
\end{eqnarray}
where $\triangle$ is the Laplacian in the flat space and $\rho$ is the matter density \cite{FOG}. Further we assumes $f_X(0)\,=\,1$ without loss of generality.

By choosing ${m_1}^2\,,{m_2}^2\,>0$ and introducing $\mu_{1,2}\,\doteq\,\sqrt{|{m_{1,2}}^2|}$ the gravitational
potential in the case of point-like source ($\rho\,=\,M\,\delta(\mathbf{x})$) is given by

\begin{eqnarray}\label{sol_pointlike}
\Phi_{pl}(\mathbf{x})\,=\,-\,\frac{GM}{|\textbf{x}|}\biggl[1+\frac{1}{3}\,e^{-\mu_1|\mathbf{x}|}-\frac{4}{3}\,
e^{-\mu_2|\mathbf{x}|}\biggr]
\end{eqnarray}
while in the case of generic matter source distribution we perform the change $\Phi\,\rightarrow\,\int d\Phi$. The passage from the pointlike source to extended one is correct only in the Newtonian limit since a such limit corresponds also to the linearized version of theory.

The motion of bodies is given by geodesic equation

\begin{eqnarray}\label{geodesic}
\frac{d^2\,x^\mu}{ds^2}+\Gamma^\mu_{\alpha\beta}\frac{dx^\alpha}{ds}\frac{dx^\beta}{ds}\,=\,0
\end{eqnarray}
where $ds\,=\,\sqrt{g_{\alpha\beta}dx^\alpha dx^\beta}$ is the relativistic distance and $\Gamma^\mu_{\alpha\beta}$ are the Christoffel symbols. In the Newtonian limit we obtain

\begin{eqnarray}
\frac{d^2\,\mathbf{x}}{dt^2}\,=\,-\nabla\Phi(\mathbf{x})
\end{eqnarray}
where the study of motion is very simple in particular cases of symmetry. For example the case of stationary motion on the circular orbit we get

\begin{eqnarray}\label{stationary_motion_2}
v_c(|\mathbf{x}|)\,=\,\sqrt{|\mathbf{x}|\,\frac{\partial\Phi(\mathbf{x})}{\partial|\mathbf{x}|}}
\end{eqnarray}

The distribution of mass can be modeled simply by introducing two sets of coordinates: the spherical coordinates $(r,\theta,\phi)$ and the cylindrical coordinates $(R,\theta,z)$. An useful mathematical tool is the Gauss flux theorem for Gravity. Since the Newtonian mechanics satisfies this theorem and, by thinking to a spherical system of mass distribution, we get, from (\ref{stationary_motion_2}), the equation

\begin{eqnarray}\label{circular_velocity}
{v_c(r)}\,=\,\sqrt{\frac{G\,M(r)}{r}}\,=\,\sqrt{\frac{4\pi G}{r}\int_0^rdy\,y^2\,\rho(y)}
\end{eqnarray}
where $M(r)$ is the only mass enclosed in the sphere with radius $r$. The Green function of the $f(X,Y,Z)$-Gravity ($\neq\,|\mathbf{x}-\mathbf{x}'|^{-1}$), instead, does not satisfy the theorem \cite{Stabile_Capozziello}. In this case we must consider directly the gravitational potential

\begin{eqnarray}\label{sol_gen}
\begin{array}{ll}
\Phi(\mathbf{x})\,=\,-\,G\int d^3\mathbf{x}'\frac{\rho(\mathbf{x}')}{|\mathbf{x}-\mathbf{x}'|}
\biggl[1+\frac{1}{3}\,e^{-\mu_1|\mathbf{x}-\mathbf{x}'|}\\\\
\qquad\qquad\qquad\qquad\qquad\qquad\qquad-\frac{4}{3}\,e^{-\mu_2|\mathbf{x}-\mathbf{x}'|}\biggr]
\end{array}
\end{eqnarray}
Apart the mathematical difficulties incoming from the research of gravitational potential for a given mass distribution, the non-validity of Gauss theorem implies, for example, that a sphere can not be reduced to a point. In fact the gravitational potential generated by a ball (also with constant density) is depending also on the Fourier transform of ball \cite{Stabile_Capozziello}. Only in the limit case where the radius of ball is small with respect to the distance we obtain the simple expression (\ref{sol_pointlike}).

We remember that in the potential (\ref{sol_gen}) we can distinguish the contributions of the bulge, the disk and the (eventual) Dark Matter. $r$ is the radial coordinate in the spherical system, while $R$, $z$ are respectively the radial coordinate in the plane of disc and the distance from the plane then we have the geometric relation $r\,=\,\sqrt{R^2+z^2}$. The main item is the choice of models of matter distribution. The more simple model characterizing the shape of galaxy is the following

\begin{eqnarray}\label{density_3}
\begin{array}{ll}
\rho_{bulge}(r)\,=\,\frac{M_b}{2\,\pi\,{\xi_b}^{3-\gamma}\,\Gamma(\frac{3-\gamma}{2})}\frac{e^{-\frac{r^2}{{\xi_b}^2}}}{r^\gamma}\\\\
\sigma_{disk}(R)\,=\,\frac{M_d}{2\pi\,{\xi_d}^2}\,
e^{-\frac{R}{\xi_d}}\\\\
\rho_{DM}(r)\,=\,\frac{\alpha\,M_{DM}}{\pi\,(4-\pi){\xi_{DM}}^3}\,\frac{1}{1+\frac{r^2}{{\xi_{DM}}^2}}
\end{array}
\end{eqnarray}
where $\Gamma(x)$ is the Gamma function, $0\,\leq\,\gamma\,<\,3$ is a free parameter and $0\,\leq\,\alpha\,<\,1$ is the ratio of Dark Matter inside the sphere with radius $\xi_{DM}$ with respect the total Dark Matter $M_{DM}$. Moreover the couples $\xi_b$, $M_b$ and $\xi_d$, $M_d$ are the radius and the mass of the bulge and the disc.

\section{Numerical analysis}\label{computation}

The computational analysis here described is referred to the study of the rotation curve (\ref{stationary_motion_2}) which can be replaced as follows
\begin{eqnarray}\label{stationary_motion_3}
v(r,R,z)=\sqrt{r\frac{\partial}{\partial r}\Phi(r,R,z)}
\end{eqnarray}
where $\Phi(r,R,z)$ is the gravitational potential
\begin{widetext}
\begin{eqnarray}\label{potential}
&&\Phi(r,R,z)\,=\,\frac{4\pi G}{3}\,\biggl[\frac{1}{r}\int_0^\infty
dr'\,\rho_{bulge}(r')\,r'\,\biggl(3\,\frac{|r-r'|-r-r'}{2}
-\frac{e^{-\mu_1|r-r'|}-e^{-\mu_1(r+r')}}{2\,\mu_1}
+2\,\frac{e^{-\mu_2|r-r'|}-e^{-\mu_2(r+r')}
}{\mu_2}\biggr)\biggr]
\nonumber\\\nonumber\\&&
+\frac{4\pi G}{3}\,\biggl[\frac{1}{r}\int_0^{\Xi}
dr'\,\rho_{DM}(r')\,r'\,\biggl(3\,\frac{|r-r'|-r-r'}{2}
-\frac{e^{-\mu_1|r-r'|}-e^{-\mu_1(r+r')}}{2\,\mu_1}
+2\,\frac{e^{-\mu_2|r-r'|}-e^{-\mu_2(r+r')}
}{\mu_2}\biggr)\biggr]
\nonumber\\\nonumber\\&&
-2\,G\,\biggr\{\int_0^\infty
dR'\,\sigma_{disc}(R')\,R'\,\biggl(\frac{\mathfrak{K}(\frac{4RR'}{(R+R')^2+z^2})}{\sqrt{(R+R')^2+z^2}}
+\frac{\mathfrak{K}(\frac{-4RR'}{(R-R')^2+z^2})}{\sqrt{(R-R')^2+z^2}}\biggr)+\int_0^\infty
dR'\,\sigma_{disc}(R')\,R'\,
\\\nonumber\\&&
\times\int_0^{\pi} d\theta'\frac{1}{3\,\sqrt{(R+R')^2+z^2-4RR'\cos^2\theta'}}
\biggl[e^{-\mu_1\sqrt{(R+R')^2+z^2-4RR'\cos^2\theta'}}
-4\,e^{-\mu_2\sqrt{(R+R')^2+z^2-4RR'\cos^2\theta'}}\biggr]\biggr\}\nonumber
\end{eqnarray}
\end{widetext}
and $\mathfrak{K}(x)$ is the Elliptic function. The parameters $\mu_1$ and $\mu_2$ are the free parameters in the theory and only by fitting process can be fixed. A sensible item is the choice of distance $\Xi$ on the which we are observing the rotation curve. In fact all models for the Dark Matter component are not limited and we need to cut the upper value of integration in (\ref{potential}).

A further distinction are the contributions to the potential coming from terms of General Relativity (GR) origin and terms of Forth Order Gravity (FOG) origin. Finally our aim is the numerical evaluation of the rotation curve in the galactic plane

\begin{eqnarray}\label{velocity}
v(R,R,0)=\sqrt{R\frac{\partial}{\partial R}\Phi(R,R,0)}
\end{eqnarray}

The first step, after the definition of the numerical values for the parameters, has been the building of the velocity starting from the derivative of the potential. For this, we need to build the definitions of the contributions to the density coming from the bulge, the disk and the dark matter (respectively \texttt{$\rho$b[r\_]}, \texttt{$\sigma$d[r\_],\texttt{$\rho$DM[r\_]}} in the full code present in appendix), together with the already mentioned splitting in the GR contributions and FOG contributions (respectively \texttt{TerGR[x\_,y\_]} and \texttt{TerYu[x\_,y\_]} in the code).

The derivative and integration operations commute, then we ``transport'' the derivative in the the integrand and then we make the integration. We found this computationally more rapid. We turning off the warning messages concerning the numerical integrations with the following commands:
\begin{quote}
\verb|Off[NIntegrate::inmur]|
\verb|Off[NIntegrate::slowcon]|
\verb|Off[NIntegrate::ncvb]|
\verb|Off[NIntegrate::eincr]|
\end{quote}
The first \verb|Off| is justified since all the variables definitions are made with the ``SetDelayed'' command (:=) that postpones the numerical evaluation of the integral making it not immediately numerical. The second turn off the message of slow convergence of the integration and making so, we avoid a long series of warning messages. The third avoid to the program to inform us of the need to use a larger number of recursive refinements in the computation. In effect, we increase the refinements with the command \verb|MaxRecursion| $\rightarrow$ \verb|20| for all the integrals, but this is not sufficient by itself to turn off the warnings. To do this, we need a bigger number than 20, but the computation became excessively slow and the final result remains practically unchanged. Since the computation is faced with oscillating error estimation, that explains the origin of the last warning message. We need to add to the numerical integration, the command \verb|Method -> GlobalAdaptive|. Still here, the warning message has to be manually closed since an improvement of \verb|Method| slows down the computation without an effective change in the results. 
An interesting thing to note, as it is possible to see in the complete code in Appendix A and there noted with the comment (*\text{$\leftarrow $}*),  is that in the definition of the derivative by means of mute variables, it need not a ``SetDelayed'' command, but a simple ``='' command, otherwise the code is unable to make the computation.

\section{Data fit}\label{data}

The next and more interesting step, is the comparison of the experimental data and what predicted by our model. From the literature cited in \citep{StSc} we can obtain the galactic speed values as function of the distance from the center and the corresponding errors. For instance, we show in some detail the manipulation of the data coming from the analysis of \citep{Stark}, concerning the external part of the Milky Way.

We start copying the data listed in the table 1 of \citep{Stark} in a table called \verb|list1|. Then we follow the prescriptions given by the authors with the introduction of new variables. As it is possible to see in the code present in Appendix B, we preserve the same notations and with the command \verb|Append| we add to the initial \verb|list1| the new variables. For instance, for the $R$ variable, with the command \verb|MapThread[Append,{list1,R}]|, we obtain a new table, here \verb|list2|, with one more column, the $R$'s valuer. And so on with the other variables. We introduce with the usual definitions, the errors on these derived quantities, here written as $\sigma x$ and added to the table. Then $\sigma$R and $\sigma\theta$ are, respectively, the error bars on the radius (the distance from the galactic center) and on the corresponding speeds and we process them together with the data in order to form the list with the values measured (or derived) and the errors on $x$ and $y$. In order to obtai a plot with the error bars, we need to load the right package for this with the command  \verb|Need["ErrorBarPlot"`]| and this make us able to to make an \verb|ErrorListPlot|. With the plot of \verb|list10|, we obtain a plot with the error bars only on y. A little bit more complicate procedure is need in order to obtain the bars on $x$ too. Indeed we need to build a list of values like \verb|{{x,y},ErrorBar{err x,err y}}| and this is done with the procedure shown in the last rows of Appendix B. In figure \ref{Stark} it is shown the result.

\begin{figure}[htbp]
  \centering
  \includegraphics[scale=0.8]{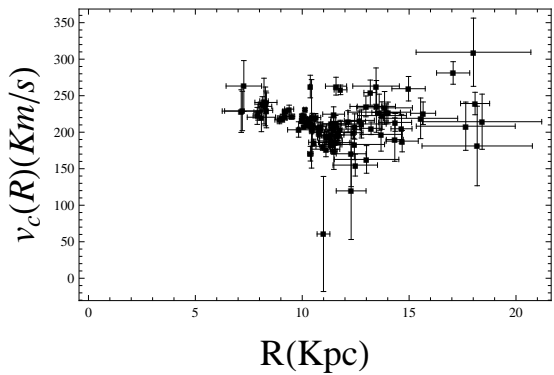}\\
  \caption{ErrorListPlot of the experimental data.}
  \label{Stark}
\end{figure}

Similar procedures for the others two part of the Milky Way data and for the NGC 3190 data. At the end, the three experimental data for the Milky Way, each corresponding to a given range of distance from the center of the galaxy, are put together in order to obtain the complete galactic rotation curve. 

At this point we proceeded following two strategies.

The first one, the faster, has been to overlap the theoretical graphs with the experimental one using the command \verb|Show|. In this case, the values of parameters in the densities (bulge, Dark Matter and disk) and of reduced masses, $\mu_1$ and $\mu_2$, are chosen by a direct overlap of the graphs.

The second strategy, more rigorous and slower, is the fit procedure. In Appendix C is present the part of code of interest. In this case, in the code of the galactic rotation curve, we fix all other parameters except the ``masses''  $\mu_1$ and $\mu_2$. These variables are the values that must be found whit the find fit procedure.

We note that the \verb|FindFit| procedure uses the parameter constraints option. In this way, it is possible to eliminate all the solutions not physically allowed and to find the values obtained by the direct investigation, that is the first strategy, $\mu_1=10^{-2}\,a^{-1}$, $\mu_2=10^{2}\,a^{-1}$ where $a$ is the characteristic scale length fixed to the value of 1 Kpc. Obviously, an increase of the number of parameters to be found with this procedure, increase he time of the computation. 

\section{Conclusions}\label{conclusions}

In this paper we present a study of the galactic rotation curve when a FOG is considered. The purely theoretical aspects have been fully exposed in \citep{StSc}. In the present work, after an obvious theoretical introduction, useful to remember the hypothesis used, we focus our attention on the salient points of the code that us permitted the computational analysis. With this program, we test the validity of our model of galactic rotation curve and the agreement  of the experimental data of two galaxies, the Milky Way and the galaxy NGC 3190, with our model.

\begin{figure}[htbp]
  \centering
  \includegraphics[scale=0.7]{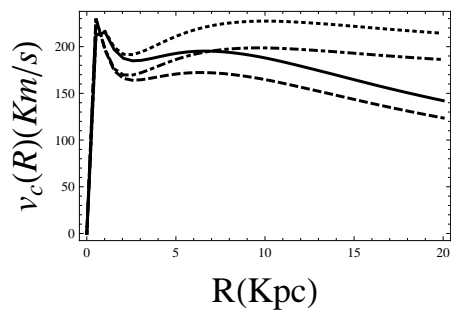}\\
    \caption{Plot of the galactic rotation curve by using the full program for Milky Way (present in Appendix A). The cases are the following: GR (dashed line), GR$+$DM (dashed and dotted line), FOG (solid line), FOG$+$DM (dotted line). The values of masses are $\mu_1\,=\,10^{-2}\,\text{Kpc}^{-1}$ and $\mu_2\,=\,10^2\,\text{Kpc}^{-1}$ \citep{StSc}}
  \label{plot_1_PRD}
\end{figure}

\begin{figure}[htbp]
  \centering
  \includegraphics[scale=0.7]{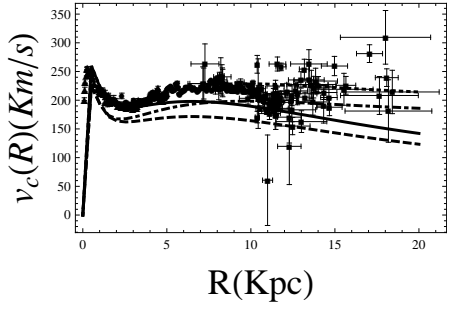}\\
  \caption{Superposition of theoretical behaviors GR (dashed line), GR$+$DM (dashed and dotted line), FOG (solid line), FOG$+$DM (dotted line) on the experimental data for Milky Way. The values of masses are $\mu_1\,=\,10^{-2}\,\text{Kpc}^{-1}$ and $\mu_2\,=\,10^2\,\text{Kpc}^{-1}$ \citep{StSc}.}
  \label{plot_2_PRD}
\end{figure}

In order to make the explanation as complete as possible and to contextualize the several pieces of code examined,  we show, in Appendix A,  the full code corresponding to the plot of the figure \ref{plot_1_PRD}, that is the code for a galaxy whose components are the bulge, the disk and the Dark Matter. The code referring also to the study of the galaxy NGC 3190 is exactly the same with the exclusion of the part of code referring to the bulge. As it is possible to see from figure \ref{plot_2_PRD}, the agreement of our model with the experimental data of the Milky Way is very good. Only for very low values of the distance $R$ the agreement is not perfect. This suggest us that we only need an improvement of the parameters in the code, maintaining the code itself essentially unchanged.

The complete code that refers to the data analysis is omitted since, apart the obvious introduction of the data coming from the cited literature, the complete program is a mere reply of what presented in Appendix B. \\

\appendix
\begin{widetext}
\section{Galactic rotation curve code}

\noindent\(\pmb{<<\text{PhysicalConstants$\grave{ }$}}\\
\pmb{\text{Gr}=\text{GravitationalConstant}[[1]];\text{MSun}=1.98\times 10^{30};}\\
\pmb{\text{$\mu $1}=10^{-2};\text{$\mu $2}=10^2;\text{Mb}=1.8\text{(*}\text{\textit{Bulge} \text{mass}}\text{*)};\text{Md}=6.5\text{(*}\text{\textit{Disk}
\text{mass}}\text{*)};\text{MDM}=4.2}\\
\pmb{\text{(*}\text{\textit{Dark} \text{Matter} \text{mass}}\text{*)};\text{$\xi $b}=0.5\text{(*}\text{\textit{Bulge} \text{radius}}\text{*)};\text{$\xi
$d}=3.5\text{(*}\text{\textit{Galaxy} \text{radius}}\text{*)};}\\
\pmb{\text{$\xi $DM}=5.5\text{(*}\text{D.M. radius}\text{*)};\chi =20;\Xi =\chi ;\text{stepr}=0.5;z=5\times 10^{-5};}\\
\pmb{\text{ab}\text{:=}\text{$\#$1} (\text{HeavisideTheta}[\text{$\#$1}]-\text{HeavisideTheta}[-\text{$\#$1}])\& }\\
\pmb{K=10^{-3}\sqrt{\frac{\text{Gr}\times 10^{10}\times \text{MSun}\times \text{Mb}}{10^3\times 3.08\times 10^{16}\times \text{$\xi $b}}};}\\
\pmb{\text{$\rho $b}[\text{r$\_$}]\text{:=}e^{-\left(\frac{r}{\text{$\xi $b}}\right)^2} \text{(*}\text{Bulge density}\text{*)}}\\
\pmb{\text{$\rho $DM}[\text{r$\_$}]\text{:=}\frac{1}{1+\left(\frac{r}{\text{$\xi $DM}}\right)^2} \text{(*}\text{Dark Matter density}\text{*)}}\\
\pmb{\text{$\sigma $d}[\text{y$\_$}]\text{:=}e^{-\frac{y}{\text{$\xi $d}}} \text{(*}\text{Disk density}\text{*)}}\\
\pmb{\text{TerGr}[\text{x$\_$},\text{y$\_$}]\text{:=}3\frac{\text{ab}[x-y]-x-y}{2}}\\
\pmb{\text{TerYu}[\text{x$\_$},\text{y$\_$}]\text{:=}-\frac{e^{- \text{$\mu $1} \text{ab}[x-y]}-e^{- \text{$\mu $1} (x+y)}}{2 \text{$\mu $1}}+2\frac{e^{-
\text{$\mu $2} \text{ab}[x-y]}-e^{-\text{$\mu $2} (x+y)}}{\text{$\mu $2}}}\\
\pmb{\text{F1}[\text{r$\_$},\text{rp$\_$}]\text{:=}\frac{\text{EllipticK}\left[\frac{4r \text{rp}}{(r+\text{rp})^2+z^2}\right]}{\sqrt{(r+\text{rp})^2+z^2}}+\frac{\text{EllipticK}\left[-\frac{4r
\text{rp}}{(r-\text{rp})^2+z^2}\right]}{\sqrt{(r-\text{rp})^2+z^2}}}\\
\pmb{\text{F2}[\text{r$\_$},\text{rp$\_$},\text{$\theta $p$\_$}]\text{:=}\frac{e^{-\text{$\mu $1} \sqrt{(r+\text{rp})^2-4 r \text{rp}\text{  }\text{Cos}[\text{$\theta
$p}]^2+z^2}}-4e^{-\text{$\mu $2} \sqrt{(r+\text{rp})^2-4 r \text{rp}\text{  }\text{Cos}[\text{$\theta $p}]^2+z^2}}}{3\sqrt{(r+\text{rp})^2-4 r \text{rp}\text{
 }\text{Cos}[\text{$\theta $p}]^2+z^2}}}\\
\pmb{\text{integrandb1}[\text{r$\_$},\text{rp$\_$},\gamma \_]\text{:=}\frac{2}{3\text{$\xi $b}^{2-\gamma }\text{Gamma}\left[\frac{3-\gamma }{2}\right]}\frac{1}{r}
\text{$\rho $b}[\text{rp}]\text{rp}^{1-\gamma }(\text{TerGr}[r,\text{rp}]+\text{TerYu}[r,\text{rp}])}\\
\pmb{\text{Derintegrandb1}[\text{r$\_$},\text{rp$\_$},\gamma \_]=D[\text{integrandb1}[r,\text{rp},\gamma ],r]; \text{(*}\text{$\leftarrow $}\text{*)}}\\
\pmb{\text{$\Phi $b1}[\text{R$\_$},\gamma \_]\text{:=}}\\
\pmb{\text{NIntegrate}[\text{Derintegrandb1}[r,\text{rp},\gamma ],\{\text{rp},0,100\text{$\xi $b}\},\text{MaxRecursion}\to 20,}\\
\pmb{\text{Method}\to \text{{``}GlobalAdaptive{''}}]\text{/.}r\to \sqrt{R^2+z^2}}\\
\pmb{\text{integrandDM1}[\text{r$\_$},\text{rp$\_$},\alpha \_]\text{:=}\frac{4 \alpha  \text{MDM}/\text{Mb}}{3(4-\pi )\left.\text{$\xi $DM}^3\right/\text{$\xi
$b}}\frac{1}{r} \text{$\rho $DM}[\text{rp}]\text{rp}(\text{TerGr}[r,\text{rp}]+\text{TerYu}[r,\text{rp}])}\\
\pmb{\text{DerintegrandDM1}[\text{r$\_$},\text{rp$\_$},\alpha \_]=D[\text{integrandDM1}[r,\text{rp},\alpha ],r];\text{(*}\text{$\leftarrow $}\text{*)}}\\
\pmb{\text{$\Phi $DM1}[\text{R$\_$},\alpha \_]\text{:=}}\\
\pmb{\text{NIntegrate}[\text{DerintegrandDM1}[r,\text{rp},\alpha ],\{\text{rp},0,\Xi \},\text{MaxRecursion}\to 20,}\\
\pmb{\text{Method}\to \text{{``}GlobalAdaptive{''}}]\text{/.}r\to \sqrt{R^2+z^2}}\\
\pmb{\text{integrandd11}[\text{r$\_$},\text{rp$\_$}]\text{:=}-\frac{(\text{Md}/\text{Mb})}{\pi \frac{\text{$\xi $d}^2}{\text{$\xi $b}}}\times (\text{rp}
\text{$\sigma $d}[\text{rp}] \text{F1}[r,\text{rp} ])}\\
\pmb{\text{integrandd12}[\text{r$\_$},\text{rp$\_$},\text{$\theta $p$\_$}]\text{:=}-\frac{(\text{Md}/\text{Mb})}{\pi \frac{\text{$\xi $d}^2}{\text{$\xi
$b}}}\times (\text{rp} \text{$\sigma $d}[\text{rp}] \text{F2}[r,\text{rp},\text{$\theta $p} ])}\\
\pmb{\text{Derintegrandd11}[\text{r$\_$},\text{rp$\_$}]=D[\text{integrandd11}[r,\text{rp}],r];}\\
\pmb{\text{Derintegrandd12}[\text{r$\_$},\text{rp$\_$},\text{$\theta $p$\_$}]=D[\text{integrandd12}[r,\text{rp},\text{$\theta $p}],r];\text{(*}\text{$\leftarrow $}\text{*)}}\\
\pmb{\text{$\Phi $d1}[\text{R$\_$}]\text{:=}}\\
\pmb{(\text{NIntegrate}[\text{Derintegrandd11}[R,\text{rp}],\{\text{rp},0,50\text{$\xi $d}\},\text{MaxRecursion}\to 20,}\\
\pmb{\text{Method}\to \text{{``}GlobalAdaptive{''}}]+}\\
\pmb{\text{NIntegrate}[\text{Derintegrandd12}[R,\text{rp},\text{$\theta $p}],\{\text{rp},0,50\text{$\xi $d}\},\{\text{$\theta $p},0,\pi \},\text{MaxRecursion}\to
20,}\\
\pmb{\text{Method}\to \text{{``}GlobalAdaptive{''}}])}\\
\pmb{\text{integrandb2}[\text{r$\_$},\text{rp$\_$},\gamma \_]\text{:=}\frac{2}{3\text{$\xi $b}^{2-\gamma }\text{Gamma}\left[\frac{3-\gamma }{2}\right]}\frac{1}{r}
\text{$\rho $b}[\text{rp}]\text{rp}^{1-\gamma }\text{TerGr}[r,\text{rp}]}\\
\pmb{\text{Derintegrandb2}[\text{r$\_$},\text{rp$\_$},\gamma \_]=D[\text{integrandb2}[r,\text{rp},\gamma ],r];\text{(*}\text{$\leftarrow $}\text{*)}}\\
\pmb{\text{$\Psi $b2}[\text{R$\_$},\gamma \_]\text{:=}}\\
\pmb{\text{NIntegrate}[\text{Derintegrandb2}[r,\text{rp},\gamma ],\{\text{rp},0,100\text{$\xi $b}\},\text{MaxRecursion}\to 20,}\\
\pmb{\text{Method}\to \text{{``}GlobalAdaptive{''}}]\text{/.}r\to \sqrt{R^2+z^2}}\\
\pmb{\text{integrandDM2}[\text{r$\_$},\text{rp$\_$},\alpha \_]\text{:=}\frac{4 \alpha  \text{MDM}/\text{Mb}}{3(4-\pi )\left.\text{$\xi $DM}^3\right/\text{$\xi
$b}}\frac{1}{r} \text{$\rho $DM}[\text{rp}]\text{rp} \text{TerGr}[r,\text{rp}]}\\
\pmb{\text{DerintegrandDM2}[\text{r$\_$},\text{rp$\_$},\alpha \_]=D[\text{integrandDM2}[r,\text{rp},\alpha ],r];\text{(*}\text{$\leftarrow $}\text{*)}}\\
\pmb{\text{$\Psi $DM2}[\text{R$\_$},\alpha \_]\text{:=}}\\
\pmb{\text{NIntegrate}[\text{DerintegrandDM2}[r,\text{rp},\alpha ],\{\text{rp},0,\Xi \},\text{MaxRecursion}\to 20,}\\
\pmb{\text{Method}\to \text{{``}GlobalAdaptive{''}}]\text{/.}r\to \sqrt{R^2+z^2}}\\
\pmb{\text{Derintegrandd2}[\text{r$\_$},\text{rp$\_$}]=D[\text{integrandd11}[r,\text{rp}],r];\text{(*}\text{$\leftarrow $}\text{*)}}\\
\pmb{\text{$\Psi $d2}[\text{R$\_$}]\text{:=}\text{NIntegrate}[\text{Derintegrandd2}[R,\text{rp}],\{\text{rp},0,50\text{$\xi $d}\},\text{MaxRecursion}\to
20,}\\
\pmb{\text{Method}\to \text{{``}GlobalAdaptive{''}}]}\\
\pmb{\Phi [\text{r$\_$},\gamma \_]\text{:=}\text{$\Phi $b1}[r,\gamma ]+\text{$\Phi $d1}[r]}\\
\pmb{\Psi [\text{r$\_$},\gamma \_]\text{:=}\text{$\Psi $b2}[r,\gamma ]+\text{$\Psi $d2}[r]}\\
\pmb{\text{$\Phi $1}[\text{r$\_$},\gamma \_,\alpha \_]\text{:=}\Phi [r,\gamma ]+\text{$\Phi $DM1}[r,\alpha ]}\\
\pmb{\text{$\Psi $1}[\text{r$\_$},\gamma \_,\alpha \_]\text{:=}\Psi [r,\gamma ]+\text{$\Psi $DM2}[r,\alpha ]}\\
\pmb{\text{VelFOG}[\text{r$\_$},\gamma \_]\text{:=}K(r \Phi [r,\gamma ]) ^{\frac{1}{2}}}\\
\pmb{\text{VelGR}[\text{r$\_$},\gamma \_]\text{:=}K(r \Psi [r,\gamma ]) ^{\frac{1}{2}}}\\
\pmb{\text{VelFOGDM}[\text{r$\_$},\gamma \_,\alpha \_]\text{:=}K(r \text{$\Phi $1}[r,\gamma ,\alpha ]) ^{\frac{1}{2}}}\\
\pmb{\text{VelGRDM}[\text{r$\_$},\gamma \_,\alpha \_]\text{:=}K(r \text{$\Psi $1}[r,\gamma ,\alpha ]) ^{\frac{1}{2}}}\\
\pmb{\text{dataVelFOG}[\gamma \_]\text{:=}\text{Table}\left[\text{VelFOG}[R,\gamma ],\left\{R,10^{-7},\chi ,\text{stepr}\right\}\right];}\\
\pmb{\text{dataVelGR}[\gamma \_]\text{:=}\text{Table}\left[\text{VelGR}[R,\gamma ],\left\{R,10^{-7},\chi , \text{stepr}\right\}\right];}\\
\pmb{\text{dataVelFOGDM}[\gamma \_,\alpha \_]\text{:=}\text{Table}\left[\text{VelFOGDM}[R,\gamma ,\alpha ],\left\{R,10^{-7},\chi , \text{stepr}\right\}\right];}\\
\pmb{\text{dataVelGRDM}[\gamma \_,\alpha \_]\text{:=}\text{Table}\left[\text{VelGRDM}[R,\gamma ,\alpha ],\left\{R,10^{-7},\chi ,\text{stepr}\right\}\right];}\\
\pmb{\text{fig1}[\gamma \_]\text{:=}\text{ListPlot}[\text{dataVelGR}[\gamma ],\text{PlotRange}\to \text{All},}\\
\pmb{\text{FrameLabel}\to \left\{\text{Style}[\text{{``}R(Kpc){''}},\text{Large},\text{Black}],\text{Style}\left[\texttt{"}v_c\text{(R)(Km/s)$\texttt{"}$},\text{Large},\text{Italic},\text{Black}\right]\right\},}\\
\pmb{\text{DataRange}\to \{0,\chi \},\text{PlotStyle}\to \text{Directive}[\text{Black},\text{Dashed},\text{Thick}],\text{Joined}\to \text{True},
}\\
\pmb{\text{AxesOrigin}\to \{0,0\},\text{Frame}\to \text{True}]}\\
\pmb{\text{fig2}[\gamma \_]\text{:=}\text{ListPlot}[\text{dataVelFOG}[\gamma ],\text{PlotRange}\to \text{All},}\\
\pmb{\text{FrameLabel}\to \left\{\text{Style}[\text{{``}R(Kpc){''}},\text{Large},\text{Black}],\text{Style}\left[\texttt{"}v_c\text{(R)(Km/s)$\texttt{"}$},\text{Large},\text{Italic},\text{Black}\right]\right\},}\\
\pmb{\text{DataRange}\to \{0,\chi \},\text{PlotStyle}\to \text{Directive}[\text{Black},\text{Thick}],\text{Joined}\to \text{True}, }\\
\pmb{\text{AxesOrigin}\to \{0,0\},\text{Frame}\to \text{True}]}\\
\pmb{\text{fig3}[\gamma \_,\alpha \_]\text{:=}\text{ListPlot}[\text{dataVelGRDM}[\gamma ,\alpha ],\text{PlotRange}\to \text{All},}\\
\pmb{\text{FrameLabel}\to \left\{\text{Style}[\text{{``}R(Kpc){''}},\text{Large},\text{Black}],\text{Style}\left[\texttt{"}v_c\text{(R)(Km/s)$\texttt{"}$},\text{Large},\text{Italic},\text{Black}\right]\right\},}\\
\pmb{\text{DataRange}\to \{0,\chi \},\text{PlotStyle}\to \text{Directive}[\text{Black},\text{DotDashed},\text{Thick}],\text{Joined}\to \text{True},
}\\
\pmb{\text{AxesOrigin}\to \{0,0\},\text{Frame}\to \text{True}]}\\
\pmb{\text{fig4}[\gamma \_,\alpha \_]\text{:=}\text{ListPlot}[\text{dataVelFOGDM}[\gamma ,\alpha ],\text{PlotRange}\to \text{All},}\\
\pmb{\text{FrameLabel}\to \left\{\text{Style}[\text{{``}R(Kpc){''}},\text{Large},\text{Black}],\text{Style}\left[\texttt{"}v_c\text{(R)(Km/s)$\texttt{"}$},\text{Large},\text{Italic},\text{Black}\right]\right\},}\\
\pmb{\text{DataRange}\to \{0,\chi \},\text{PlotStyle}\to \text{Directive}[\text{Black},\text{Dotted},\text{Thick}],\text{Joined}\to \text{True},
}\\
\pmb{\text{AxesOrigin}\to \{0,0\},\text{Frame}\to \text{True}]}\\
\pmb{\alpha =0.5;\gamma =1.5;}\\
\pmb{\text{th}=\text{Show}[\text{fig1}[\gamma ],\text{fig2}[\gamma ],\text{fig3}[\gamma ,\alpha ],\text{fig4}[\gamma ,\alpha ]]}\)
\section{Data analysis code}
\noindent\(\pmb{\text{R0}=10;\text{$\omega $0}=220/\text{R0};\text{$\sigma $l}=0.05;}\\
\pmb{R=\left(\text{R0}^2+ \text{list1}[[\text{All},3]]^2-2\text{R0} \text{list1}[[\text{All},3]]\text{Cos}[\text{list1}[[\text{All},1]]{}^{\circ}]\right)^{\frac{1}{2}};}\\
\pmb{\text{list2}=\text{MapThread}[\text{Append},\{\text{list1},R\}];}\\
\pmb{\omega =\frac{\text{list2}[[\text{All},5]]}{\text{R0} \text{Sin}[\text{list2}[[\text{All},1]]{}^{\circ}]\text{Cos}[\text{list2}[[\text{All},2]]{}^{\circ}]}+\text{$\omega
$0}-}\\
\pmb{\frac{4.2 \text{Cos}[\text{list2}[[\text{All},1]]{}^{\circ}]}{\text{R0} \text{Sin}[\text{list2}[[\text{All},1]]{}^{\circ}]};}\\
\pmb{\text{list3}=\text{MapThread}[\text{Append},\{\text{list2},\omega \}];}\\
\pmb{\sigma \omega =\text{Abs}\left[(\omega -\text{$\omega $0})\sqrt{\left(\frac{\text{list3}[[\text{All},6]]}{\text{list3}[[\text{All},5]]}\right)^2+\left(\frac{\text{$\sigma
$l}}{\text{Tan}[\text{list3}[[\text{All},1]]{}^{\circ}]}\right)^2}\right];}\\
\pmb{\text{list4}=\text{MapThread}[\text{Append},\{\text{list3},\sigma \omega \}];}\\
\pmb{\Theta =R\times \omega ;}\\
\pmb{\text{list5}=\text{MapThread}[\text{Append},\{\text{list4},\Theta \}];}\\
\pmb{\text{$\sigma $R}=}\\
\pmb{\text{Abs}[}\\
\pmb{\frac{1}{R}\surd \left((\text{list4}[[\text{All},3]]-\text{R0} \text{Cos}[\text{list4}[[\text{All},1]]{}^{\circ}])^2\text{list4}[[\text{All},4]]^2+\right.}\\
\pmb{\left.\left.(\text{R0} \text{list4}[[\text{All},3]]\text{Sin}[\text{list4}[[\text{All},1]]{}^{\circ}])^2\text{$\sigma $l}^2\right)\right];}\\
\pmb{\text{list6}=\text{MapThread}[\text{Append},\{\text{list5},\text{$\sigma $R}\}];}\\
\pmb{\text{list7}=\text{Drop}[\text{list6},\{\},\{1,6\}];}\\
\pmb{\sigma \Theta =\text{Abs}\left[\Theta \sqrt{\left(\frac{\sigma \omega }{\omega }\right)^2+\left(\frac{\text{$\sigma $R}}{R}\right)^2}\right];}\\
\pmb{\text{list8}=\text{MapThread}[\text{Append},\{\text{list7},\sigma \Theta \}];}\\
\pmb{\text{list9}=\text{Drop}[\text{list8},\{\},\{2,3\}]\text{(*Errors on x and y*)};}\\
\pmb{\text{list10}=\text{Drop}[\text{list9},\{\},\{3\}]\text{(*Errors on y*)};}\\
\pmb{\text{Needs}[\text{{``}ErrorBarPlots$\grave{ }${''}}]}\\
\pmb{\text{ErrorListPlot}[\text{list10},}\\
\pmb{\text{AxesLabel}\to \{\text{Style}[\text{{``}R(Kpc){''}},\text{Large},\text{Black}],}\\
\pmb{\left.\text{Style}\left[\texttt{"}v_c\text{(R)(Km/s)$\texttt{"}$},\text{Large},\text{Italic},\text{Black}\right]\right\},\text{PlotStyle}\to
\text{Black},}\\
\pmb{\text{PlotMarkers}\to \{\text{{``}$\blacksquare${''}}\}, \text{PlotRange}\to \text{All},\text{AxesOrigin}\to \{0,0\}]}\) \text{(*}\text{Error bars only on y}\text{*)}\\
\\
\\
\noindent\(\pmb{\text{data}=\text{Drop}[\text{list9},\{\},\{3,4\}]}\\
\pmb{\text{errx}=\text{Drop}[\text{Drop}[\text{list9},\{\},\{1,2\}],\{\},\{2\}][[\text{All},1]]}\\
\pmb{\text{erry}=\text{Drop}[\text{Drop}[\text{list9},\{\},\{1,2\}],\{\},\{1\}][[\text{All},1]]}\\
\pmb{\text{listerr}=\{\text{errx},\text{erry}\}^{\mathsf{T}}}\\
\pmb{\text{error}=\text{Cases}[\text{listerr},\{\text{x$\_$},\text{y$\_$}\}:\to \text{ErrorBar}[x,y]]}\\
\pmb{\text{valerr}=\text{Map}[\{\#[[1]],\#[[2]]\}\&,\{\text{data},\text{error}\}^{\mathsf{T}}]}\\
\pmb{\text{ErrorListPlot}[\text{valerr},}\\
\pmb{\text{FrameLabel}\to \{\text{Style}[\text{{``}R(Kpc){''}},\text{Large},\text{Black}],}\\
\pmb{\left.\text{Style}\left[\texttt{"}v_c\text{(R)(Km/s)$\texttt{"}$},\text{Large},\text{Italic},\text{Black}\right]\right\},\text{PlotStyle}\to
\text{Black},}\\
\pmb{\text{PlotMarkers}\to \{\text{{``}$\blacksquare${''}}\}, \text{PlotRange}\to \text{All},\text{AxesOrigin}\to \{0,0\},\text{Frame}\to \text{True}]}\)\text{(*}\text{Error bars on x and y}\text{*)}\\
\section{Find Fit}
\noindent\(\pmb{\text{$\Phi $1}[\text{R$\_$},\text{$\mu $1$\_$},\text{$\mu $2$\_$}]\text{:=}(\text{$\Phi $b1}[R,\text{$\mu $1},\text{$\mu $2}]+\text{$\Phi
$d1}[R,\text{$\mu $1},\text{$\mu $2}]+\text{$\Phi $DM1}[R,\text{$\mu $1},\text{$\mu $2}])}\\
\pmb{\text{VelFOGDM}[\text{R$\_$},\text{$\mu $1$\_$},\text{$\mu $2$\_$}]\text{:=}K(R \text{$\Phi $1}[R,\gamma ,\text{$\mu $1},\text{$\mu $2}]) ^{\frac{1}{2}}}\\
\pmb{\text{model}= \text{VelFOGDM}[R,\text{$\mu $1},\text{$\mu $2}];}\\
\pmb{\text{FindFit}\left[\text{val},\text{model},\left\{\left\{\text{$\mu $1},10^{-2}\right\},\left\{\text{$\mu $2},10^2\right\}\right\},R\right]}\)
\end{widetext}

\end{document}